\input{aipcheck}

\documentclass[
    ,final            
  ]
  {aipproc}

\layoutstyle{6x9}

\begin{document}

\title{On the effects of rotation during the formation of population III protostars}


\author{Jayanta Dutta}{
  address={Institut f\"ur Theoretische  Astrophysik, Zentrum f\"ur Astronomie der Universit\"at Heidelberg, Albert-Ueberle-Str.\ 2, 69120 Heidelberg, Germany}
}

\author{Paul C. Clark}{
  address={Institut f\"ur Theoretische  Astrophysik, Zentrum f\"ur Astronomie der Universit\"at Heidelberg, Albert-Ueberle-Str.\ 2, 69120 Heidelberg, Germany}
}

\author{Ralf S. Klessen}{
  address={Institut f\"ur Theoretische  Astrophysik, Zentrum f\"ur Astronomie der Universit\"at Heidelberg, Albert-Ueberle-Str.\ 2, 69120 Heidelberg, Germany}
}

\begin{abstract}

It has been suggested that turbulent motions are responsible for the transport of 
angular momentum during the formation of Pop. III stars, however the exact details 
of this process have never been studied. We report the results from three dimensional 
SPH simulations of a rotating self-gravitating primordial molecular cloud, in which 
the initial velocity of solid-body rotation has been changed. We also examine the 
build-up of the discs that form in these idealized calculations.

\end{abstract}

\maketitle


\section{Model Parameters}

The Pop.\ III simulations start with a number density of $10^{3}$ cm$^{-3}$, 
with a temperature of 250 K and contain 2 Jeans masses with total mass 
$M \sim$ 2982 M$_{\odot}$ of gas.
Our clouds are given different levels of rotational support, 
described via the $\beta$-parameter defined as \\

$\beta = \frac {Rotational ~Energy} {Gravitational~ Energy} = \frac{R_0^3\Omega^2}{3GM}
=$ 0.005, 0.01, 0.05, 0.1, 0.2, \\

where $\Omega$ is the angular velocity. The clouds are all modeled 
using 5,000,000 SPH particles. The initial density($\rho$) is uniform, with the 
$t_{ff}=$1.37 Myr and $t_{sound}=$ 5 Myr.

\section{Power-Law Angular Momentum Profile}


\begin{figure*}
  \includegraphics[height=.5\textheight]{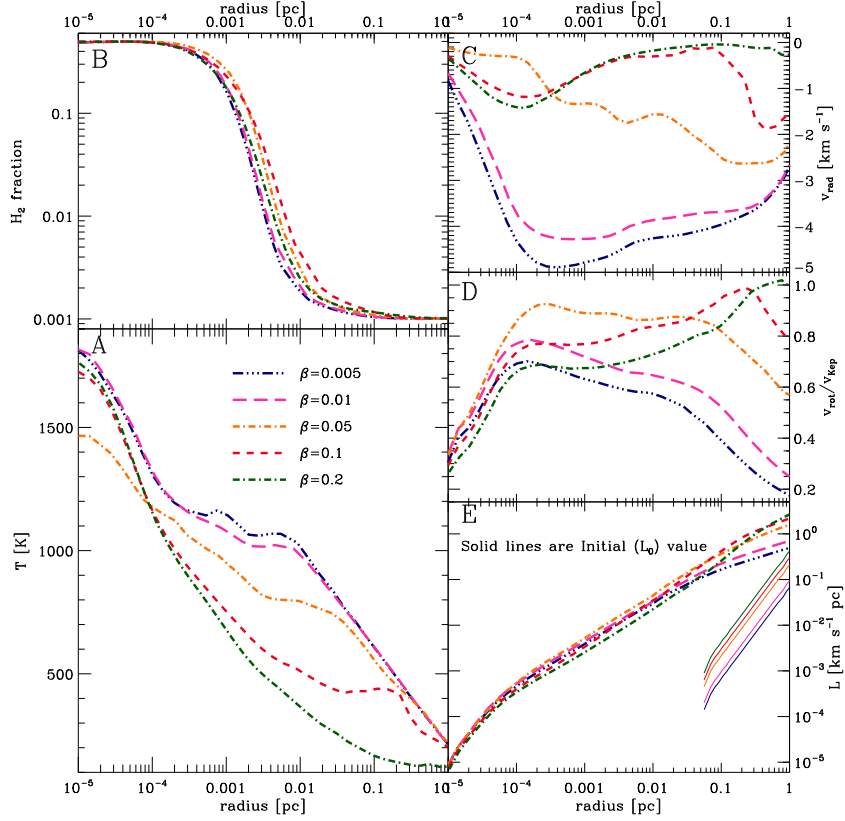}
  \caption{Radially binned, mass-weighted averages of physical 
quantities related to the angular momentum of the gas for five different $\beta$, 
compared when the gas density reaches $\sim$ 5$\times10^{-10}$ g $cm^{-3}$.}
\end{figure*}

\begin{itemize}

\item Similar H$_2$-fractions for different $\beta$ suggest that the
collapsed clouds subject to compressional heating rather than just the H$_2$ 
formation heating. A lower rate of compressional heating implies a lower radial 
velocity, and is consistent with higher degree of rotational support 
\cite{abn02},\cite{cgsgkb11}. 
 
\item  Even though our initial conditions lack any turbulence and are
very different from fully self-consistent cosmological initial conditions, we 
find the characteristic power-law angular momentum($L$) profile that has been 
reported in 
several studies, suggesting {\it universal} collapse property.
$L_0 \propto r^2$ (initially) and $L \propto r$ (power-law).

\item The origin of the slope is just the cloud getting rid of excess angular 
momentum while trying to collapse (with $\rho(r) \propto r^{-2.2}$) and thus 
dragging the remaining angular momentum along for the ride.
\end{itemize}

\section{Relative Strength of Torques}
 Gravitational torque is always less than the hydro torque. The non-axisymmetric 
nature of the cloud and the spiral arms are caused by the gravitational
instability \cite{larson84}. Yet the resulting torques are not dominated by the 
gravitational force itself, but rather the non-axisymmetric pressure
forces that arise from the structure. Hence transport is {\it local} \cite{lr04},
even though it is induced by a global phenomenon. In the outer regime the transfer 
of angular momentum happens because of particle's infall radially due to gravity.
The Keplerian radius is larger for faster-rotating clouds. The peaks in the ratio 
represent the formation of consecutive discs inside the initial disc.

\begin{figure}
  \includegraphics[height=.2\textheight]{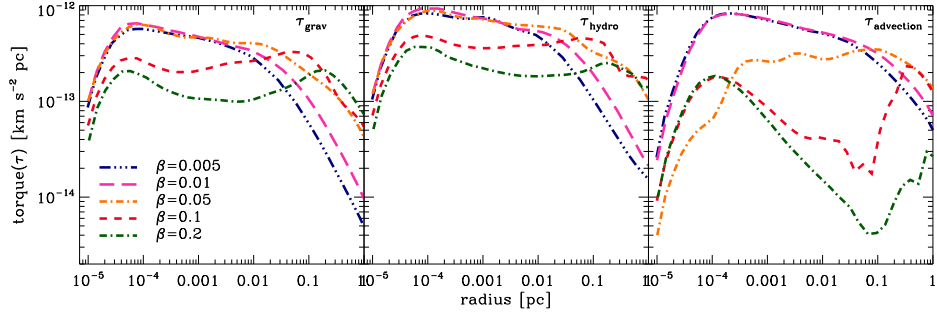}
  \caption{The gravitational, hydrodynamical and advectional torque during star formation process.}
\end{figure}

\begin{figure}
  \includegraphics[height=.3\textheight]{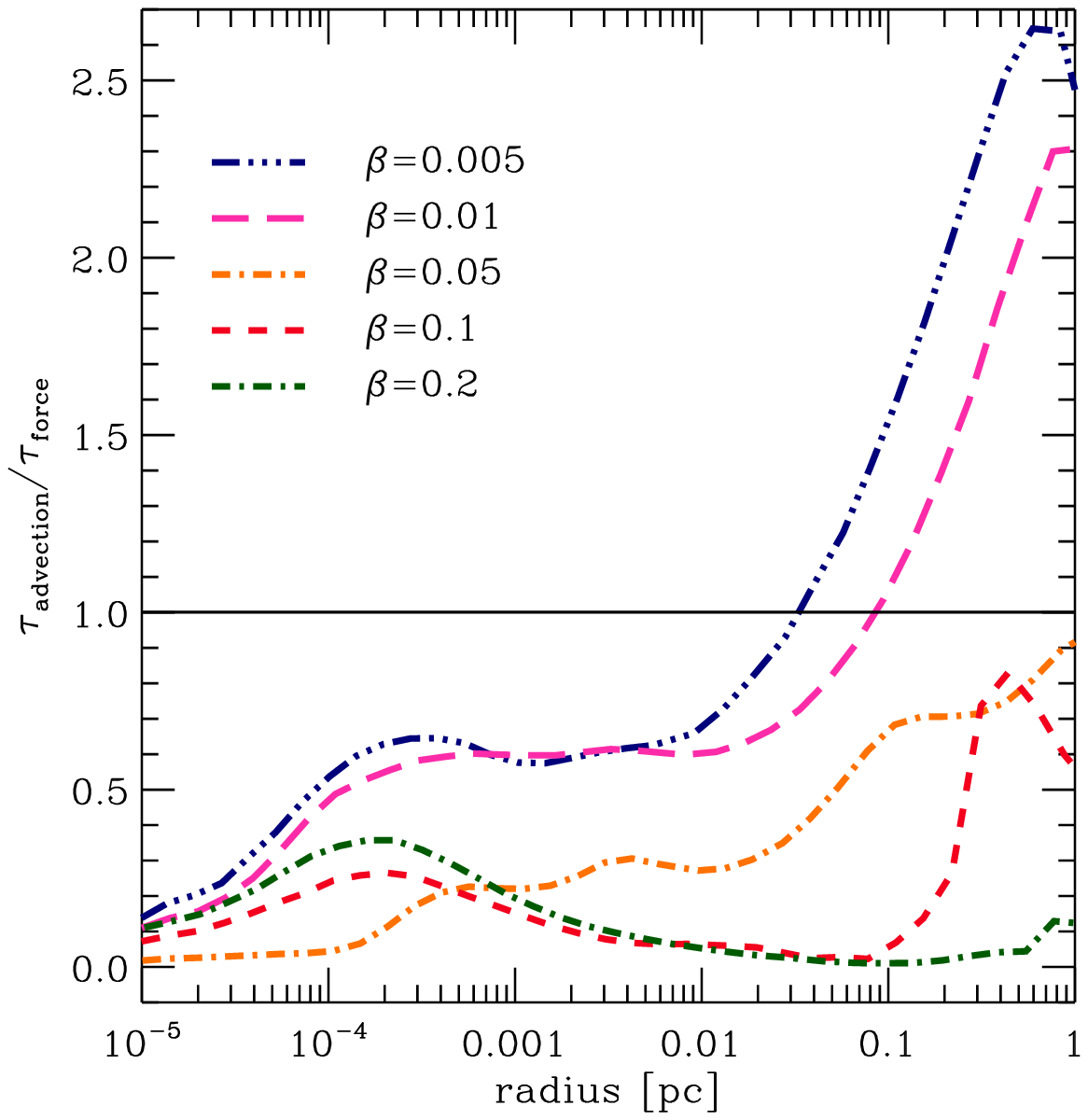}
\includegraphics[height=.3\textheight]{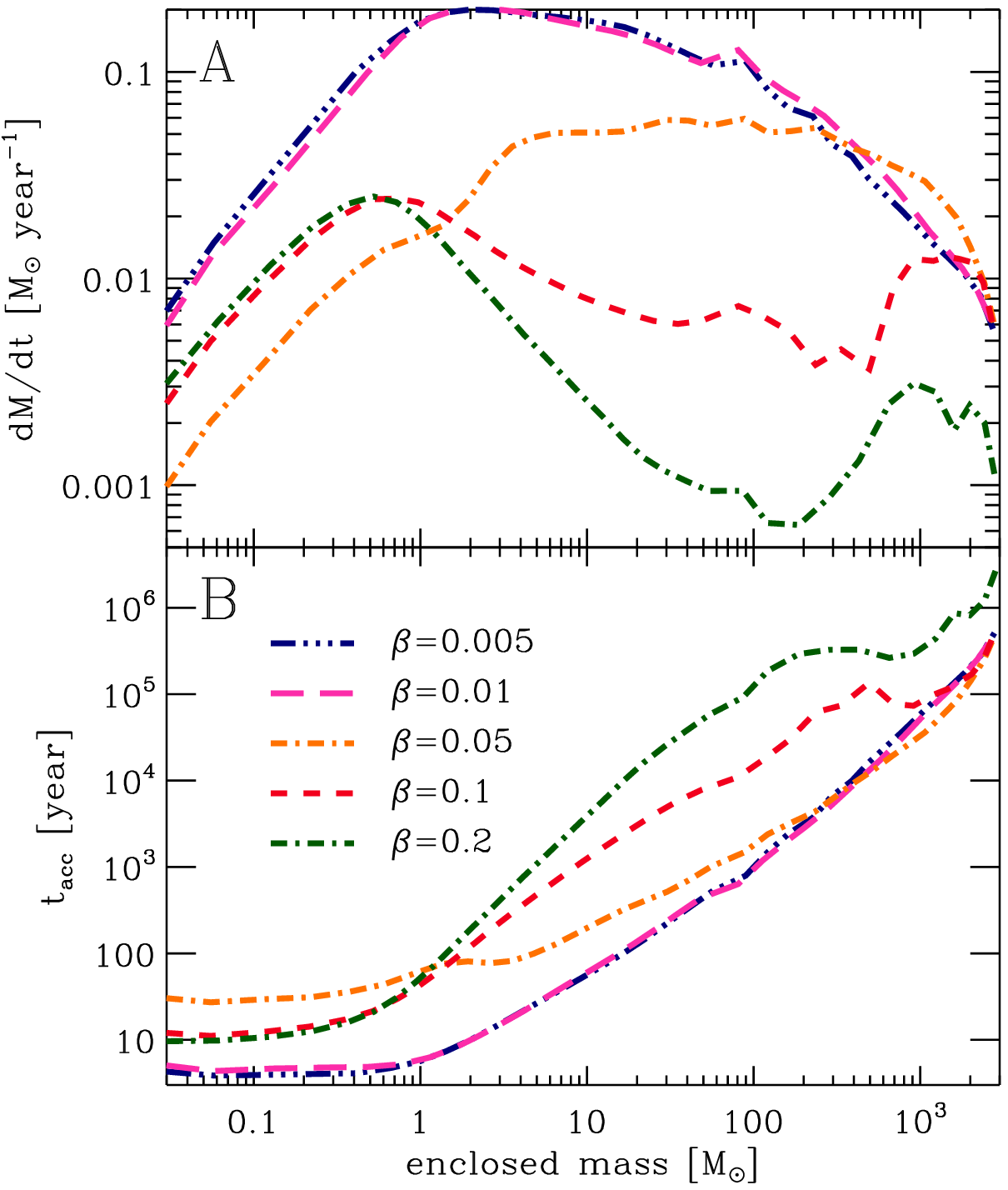}
  \caption{Relative strength of torques ({\it left}), accretion rate and accretion time ({\it right}).}
\end{figure}

\section*{Accretion Rate}

Although the time taken to accrete the inner 1M$_{\odot}$ differs only by a 
factor of 10, we see that the accretion timescales for the remainder of the 
cloud can differ substantially, depending on the level of initial rotation. 
When we consider that the number of protostars also varies substantially 
with the spin of the cloud, we would expect the luminosity from protostellar 
population will also be a strong function of the cloud's rotational properties.

\section*{Disc Fragmentation}
\begin{figure}
  \includegraphics[height=.4\textheight]{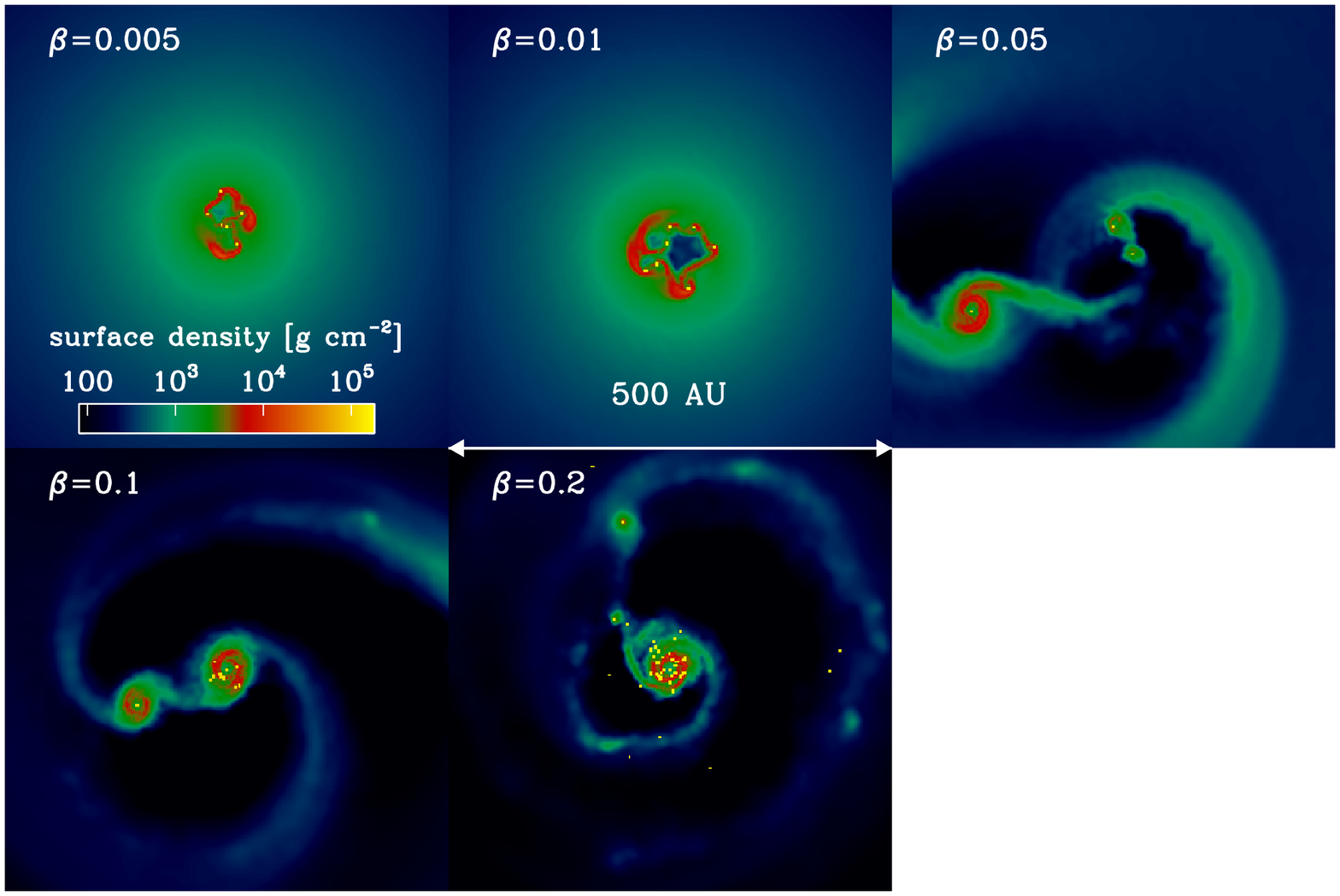}
  \caption{Surface density in 500 AU around first protostar as the young stellar system gains mass.}
\end{figure}

\begin{figure}
  \includegraphics[height=.21\textheight]{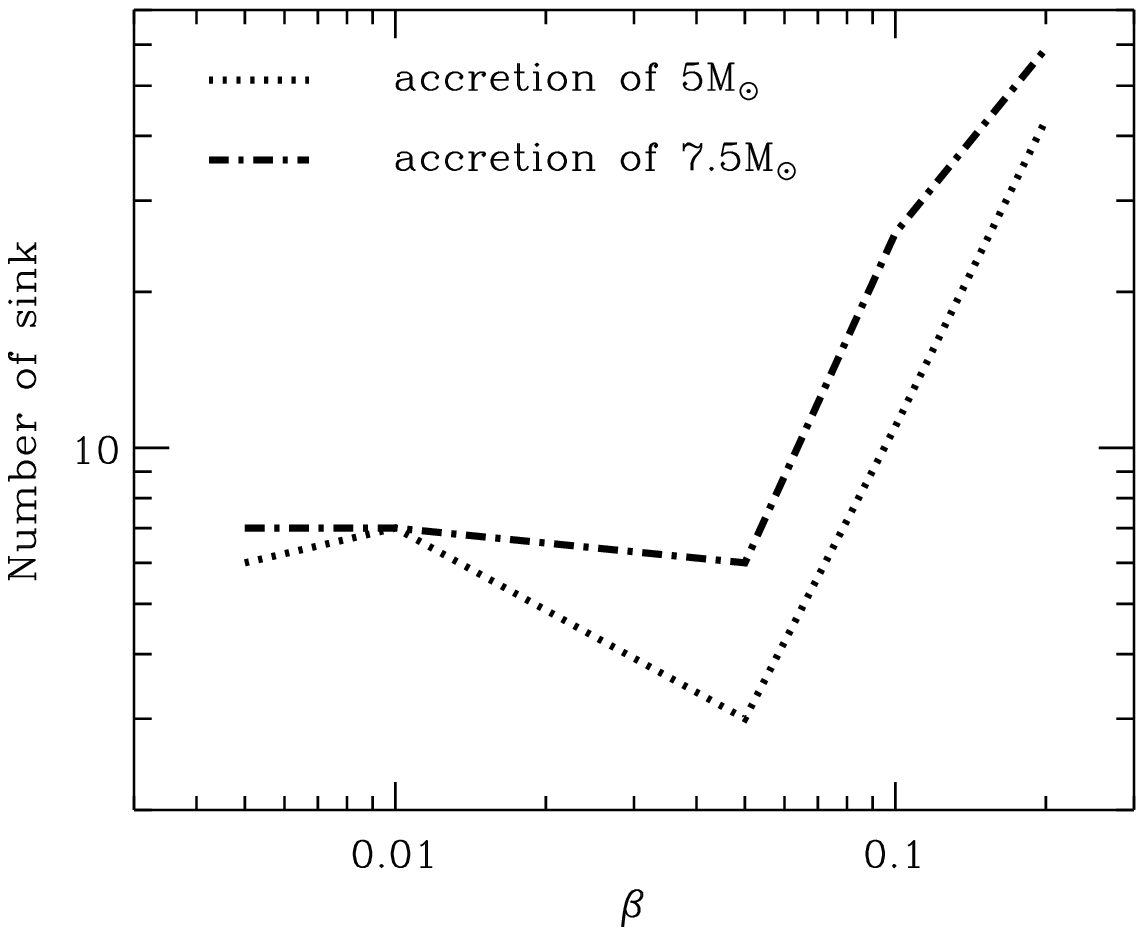}
\includegraphics[height=.21\textheight]{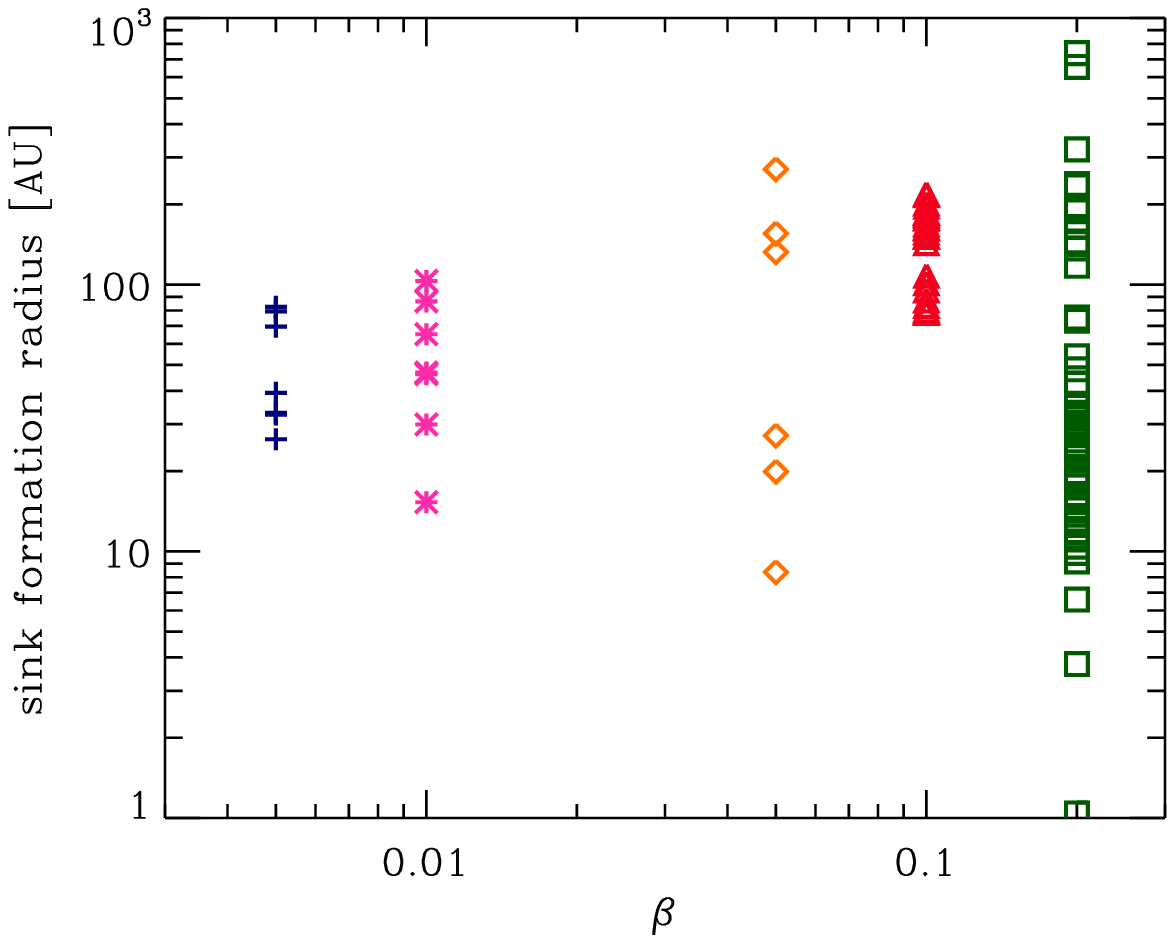}
\includegraphics[height=.21\textheight]{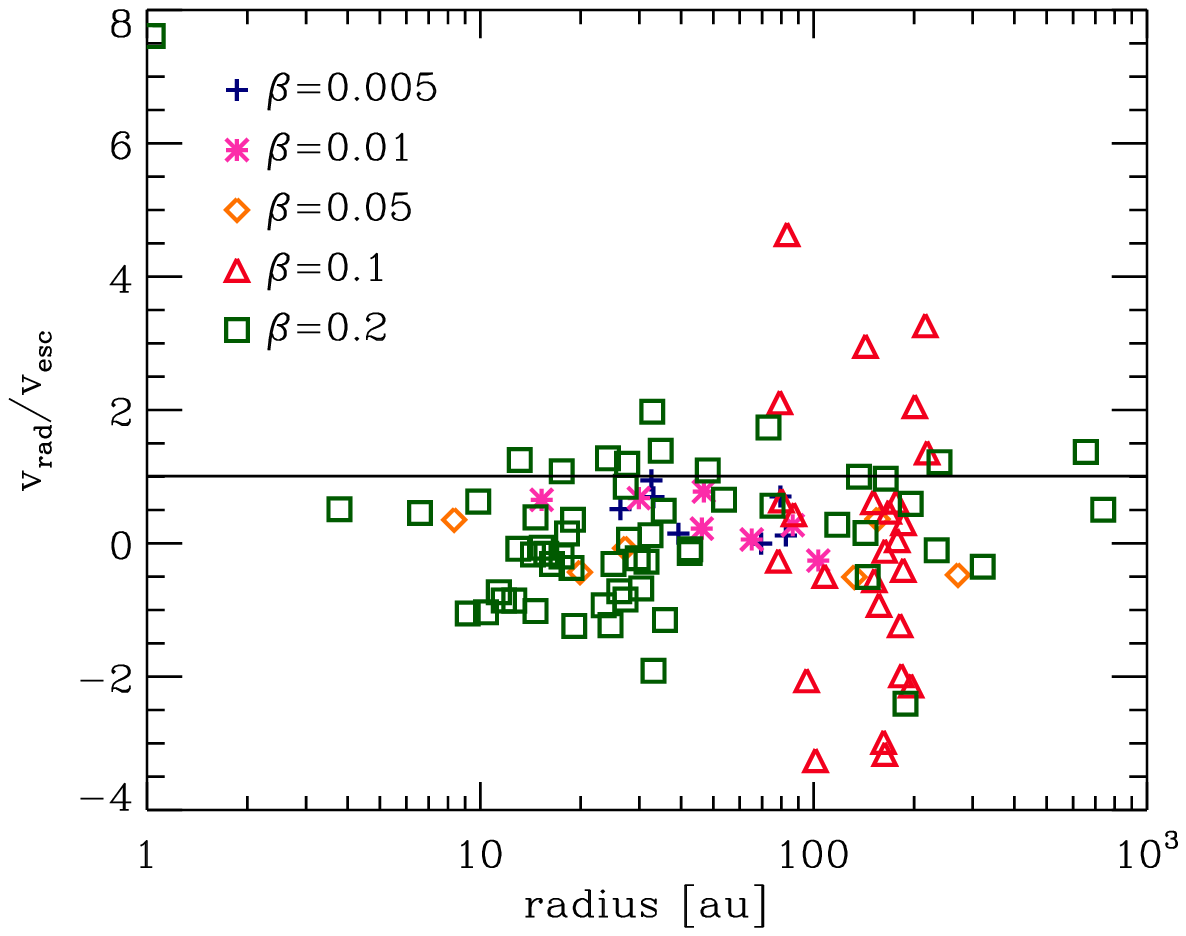}
  \caption{Number of sink particles and their escape velocities during the disc fragmentation.}
\end{figure}

Faster-rotating clouds produce more sinks as they form rotationally supported discs 
much earlier than slowly-rotating clouds. A number of protostars move away 
from their cluster with velocity exceeding escape velocity. Though the resolution
used in \cite{gswgcskb11} was much higher than our resolution, we still see
ejection from the cluster. Lower $\beta$ remains in the cluster. The range of 
sink-formation-radius increases with $\beta$, except for $\beta$ = 0.1.  

\section{Conclusions}

The angular momentum profile follows a power-law, regardless of initial rotation in the 
cloud. The transport is {\it local} and happens due to torques. The mass accretion rate 
depends on the initial level of rotation. No protostars with $\beta$ = 0.1 form within 
the central 100 AU. The faster rotating clouds escape from the cluster, might stop
accreting and enter the main sequence as low mass Pop. III stars.




\begin{theacknowledgments}
The author is thankful to Prof. R. Larson and Prof. V. Bromm for useful comments.  \end{theacknowledgments}



\bibliographystyle{aipproc}   

\bibliography{dutta}

\IfFileExists{\jobname.bbl}{}
 {\typeout{}
  \typeout{******************************************}
  \typeout{** Please run "bibtex \jobname" to optain}
  \typeout{** the bibliography and then re-run LaTeX}
  \typeout{** twice to fix the references!}
  \typeout{******************************************}
  \typeout{}
 }

\end{document}